\input harvmac
\input epsf
\input amssym
\baselineskip 14pt

\newcount\figno
\figno=0
\def\fig#1#2#3{
\par\begingroup\parindent=0pt\leftskip=1cm\rightskip=1cm\parindent=0pt
\baselineskip=11pt
\midinsert
\epsfxsize=#3
\centerline{\epsfbox{#2}}
\vskip -21pt
{\bf Fig.\ \the\figno: } #1\par
\endinsert\endgroup\par
}
\def\figlabel#1{\xdef#1{\the\figno}}
\def\encadremath#1{\vbox{\hrule\hbox{\vrule\kern8pt\vbox{\kern8pt
\hbox{$\displaystyle #1$}\kern8pt}
\kern8pt\vrule}\hrule}}

\def\p{\partial}

\def\rt{\rightarrow}

\def\zb{\overline{z}}

\def\eps{\epsilon}

\def\zb{\overline{z}}

\def\ab{\overline{a}}

\def\Tb{\overline{T}}

\def\cb{\overline{c}}

\def\Tb{\overline{T}}

\def\rh{\hat{r}}

\def\Tb{\overline{T}}

\def\Ec{{\cal E}}

\lref\McGoughLOL{
  L.~McGough, M.~Mezei and H.~Verlinde,
  ``Moving the CFT into the bulk with $T\bar T$,''
[arXiv:1611.03470 [hep-th]].
}

\lref\FaulknerJY{
  T.~Faulkner, H.~Liu and M.~Rangamani,
  ``Integrating out geometry: Holographic Wilsonian RG and the membrane paradigm,''
JHEP {\bf 1108}, 051 (2011).
[arXiv:1010.4036 [hep-th]].
}

\lref\HeemskerkHK{
  I.~Heemskerk and J.~Polchinski,
  ``Holographic and Wilsonian Renormalization Groups,''
JHEP {\bf 1106}, 031 (2011).
[arXiv:1010.1264 [hep-th]].
}

\lref\HH{
  H. Headrick and V. Hubeny, to appear.
}

\lref\AndradeFNA{
  T.~Andrade and D.~Marolf,
  ``Asymptotic Symmetries from finite boxes,''
Class.\ Quant.\ Grav.\  {\bf 33}, no. 1, 015013 (2016).
[arXiv:1508.02515 [gr-qc]].
}

\lref\ZamolodchikovCE{
  A.~B.~Zamolodchikov,
  ``Expectation value of composite field T anti-T in two-dimensional quantum field theory,''
[hep-th/0401146].
}

\lref\AharonyIPA{
  O.~Aharony and Z.~Komargodski,
  ``The Effective Theory of Long Strings,''
JHEP {\bf 1305}, 118 (2013).
[arXiv:1302.6257 [hep-th]].
}

\lref\CavagliaODA{
  A.~Cavaglià, S.~Negro, I.~M.~Szécsényi and R.~Tateo,
  ``$T \bar{T}$-deformed 2D Quantum Field Theories,''
JHEP {\bf 1610}, 112 (2016).
[arXiv:1608.05534 [hep-th]].
}

\lref\SmirnovLQW{
  F.~A.~Smirnov and A.~B.~Zamolodchikov,
  ``On space of integrable quantum field theories,''
Nucl.\ Phys.\ B {\bf 915}, 363 (2017).
[arXiv:1608.05499 [hep-th]].
}

\lref\GiveonNIE{
  A.~Giveon, N.~Itzhaki and D.~Kutasov,
  ``$ T\Tb $ and LST,''
JHEP {\bf 1707}, 122 (2017).
[arXiv:1701.05576 [hep-th]].
}

\lref\DubovskyCNJ{
  S.~Dubovsky, V.~Gorbenko and M.~Mirbabayi,
  ``Asymptotic fragility, near AdS$_{2}$ holography and $ T\overline{T} $,''
JHEP {\bf 1709}, 136 (2017).
[arXiv:1706.06604 [hep-th]].
}

\lref\GiribetIMM{
  G.~Giribet,
  ``$T\bar{T}$-deformations, AdS/CFT and correlation functions,''
[arXiv:1711.02716 [hep-th]].
}

\lref\ShyamZNQ{
  V.~Shyam,
  ``Background independent holographic dual to $T\bar{T}$ deformed CFT with large central charge in 2 dimensions,''
JHEP {\bf 1710}, 108 (2017).
[arXiv:1707.08118 [hep-th]].
}

\lref\GuicaLIA{
  M.~Guica,
  ``An integrable Lorentz-breaking deformation of two-dimensional CFTs,''
[arXiv:1710.08415 [hep-th]].
}

\lref\CollinsNJ{
  J.~C.~Collins, A.~V.~Manohar and M.~B.~Wise,
  ``Renormalization of the vector current in QED,''
Phys.\ Rev.\ D {\bf 73}, 105019 (2006).
[hep-th/0512187].
}

\lref\BalasubramanianJD{
  V.~Balasubramanian and P.~Kraus,
  ``Space-time and the holographic renormalization group,''
Phys.\ Rev.\ Lett.\  {\bf 83}, 3605 (1999).
[hep-th/9903190].
}

\lref\deBoerTGO{
  J.~de Boer, E.~P.~Verlinde and H.~L.~Verlinde,
  ``On the holographic renormalization group,''
JHEP {\bf 0008}, 003 (2000).
[hep-th/9912012].
}

\lref\BrattanMY{
  D.~Brattan, J.~Camps, R.~Loganayagam and M.~Rangamani,
  ``CFT dual of the AdS Dirichlet problem : Fluid/Gravity on cut-off surfaces,''
JHEP {\bf 1112}, 090 (2011).
[arXiv:1106.2577 [hep-th]].
}

\lref\MarolfDR{
  D.~Marolf and M.~Rangamani,
  ``Causality and the AdS Dirichlet problem,''
JHEP {\bf 1204}, 035 (2012).
[arXiv:1201.1233 [hep-th]].
}

\lref\SmirnovLQW{
  F.~A.~Smirnov and A.~B.~Zamolodchikov,
  ``On space of integrable quantum field theories,''
Nucl.\ Phys.\ B {\bf 915}, 363 (2017).
[arXiv:1608.05499 [hep-th]].
}

\lref\CavagliaODA{
  A.~Cavaglià, S.~Negro, I.~M.~Szécsényi and R.~Tateo,
  ``$T \bar{T}$-deformed 2D Quantum Field Theories,''
JHEP {\bf 1610}, 112 (2016).
[arXiv:1608.05534 [hep-th]].
}

\lref\BrownGS{
  J.~D.~Brown, J.~Creighton and R.~B.~Mann,
  ``Temperature, energy and heat capacity of asymptotically anti-de Sitter black holes,''
Phys.\ Rev.\ D {\bf 50}, 6394 (1994).
[gr-qc/9405007].
}

\lref\BrownBR{
  J.~D.~Brown and J.~W.~York, Jr.,
  ``Quasilocal energy and conserved charges derived from the gravitational action,''
Phys.\ Rev.\ D {\bf 47}, 1407 (1993).
[gr-qc/9209012].
}

\lref\BalasubramanianRE{
  V.~Balasubramanian and P.~Kraus,
  ``A Stress tensor for Anti-de Sitter gravity,''
Commun.\ Math.\ Phys.\  {\bf 208}, 413 (1999).
[hep-th/9902121].
}

\lref\coleman{
S. Coleman, ``Aspects of Symmetry", Cambridge University Press, 1985. }



\Title{\vbox{\baselineskip14pt
\hbox{CALT-TH-2018-002}
}} {\vbox{\centerline {Cutoff AdS$_3$ versus the $T\bar{T}$ deformation}}}
\centerline{Per Kraus$^1$, Junyu Liu$^2$, Donald Marolf$^3$}
\bigskip
\centerline{\it{$^1$Mani L. Bhaumik Institute for Theoretical Physics}}
\centerline{${}$\it{University of California, Los Angeles, CA 90095, USA}}
\centerline{\it{$^2$Walter Burke Institute for Theoretical Physics}}
\centerline{${}$\it{California Institute of Technology, Pasadena, CA 91125, USA}}
\centerline{\it{$^3$Department of Physics}}
\centerline{${}$\it{University of California, Santa Barbara, CA 93106, USA}}

\baselineskip14pt

\vskip .3in

\centerline{\bf Abstract}
\vskip.2cm

\noindent

A recent proposal relates two dimensional holographic conformal field theories deformed by the integrable $T\Tb$ flow to AdS$_3$ with a finite radial cutoff. We investigate this proposal by studying perturbative correlation functions on the two sides.  For low point correlators of the stress tensor, we successfully match the deformed CFT results at large central charge  to bulk results obtained in classical pure gravity.  The deformed CFT also provides definite predictions for loop corrections in the bulk. We then include matter fields in the bulk.  To reproduce the classical bulk two-point function of a scalar operator we show that the  deformed CFT needs to be augmented with double trace scalar operators, with the $T\Tb$ operator yielding corrections corresponding to loops in the bulk.

\Date{January  2018}
\baselineskip14pt

\newsec{Introduction}

Recently, McGough, Mezei, and Verlinde \McGoughLOL\  proposed an intriguing extension of the AdS$_3$/CFT$_2$ correspondence.  On the bulk side, the boundary lies not at asymptotic infinity, but instead at a finite radial position.  The dual quantum field theory is no longer conformal, but is rather described by a CFT deformed by the remarkable $T\Tb$ operator of Zamolodchikov \ZamolodchikovCE.  The bulk side of this proposed duality is interesting in that the ability to move the boundary inward could shed on the important question of the emergence of bulk locality; the notion of introducing a cutoff boundary surface in this context has arisen in earlier work, e.g.,  \refs{\BalasubramanianJD,\deBoerTGO}, and also in relation to the fluid-gravity correspondence, e.g.  \BrattanMY. In particular, \HeemskerkHK\ and (more explicitly) \FaulknerJY\ both show that such cutoffs are dual to {\it some} deformation of the orginal CFT.  Recent  work on this and related duality proposals include \refs{ \GiveonNIE,\DubovskyCNJ,\ShyamZNQ,\GuicaLIA,\GiribetIMM}.

The particular deformation proposed by \McGoughLOL\ is especially interesting: the $T\Tb$ operator is irrelevant in the renormalization group sense, yet the deformed theory appears to be far more predictive than the generic non-renormalizable QFT.  For these reasons, it is worthwhile to see to what extent the setup of \McGoughLOL\ can be elevated to a full-fledged holographic correspondence, complete with a well defined dictionary for relating observables on the two sides, and this is the focus of the present work.

The proposal of \McGoughLOL\ was not so much derived as motivated based on observing a nontrivial correspondence between several quantities computed on the two sides, in particular involving the deformed energy spectrum of certain states  and the propagation speeds of small perturbations around thermal states. Let us first review some key aspects of the $T\Tb$ deformation in QFT.  Given any 2d QFT with a local stress tensor, the composite operator $T\Tb$ can be defined in a canonical way up to derivatives of local operators \ZamolodchikovCE.  The deformed action, $S(\lambda)$, is stipulated to obey ${dS\over d\lambda} = \int\! d^2x \sqrt{g} T\Tb$.  Assuming that the undeformed theory is a CFT, as reviewed below we can equivalently say that the trace of the deformed stress tensor obeys (up to derivatives of local operators)
\eqn\aa{ T^i_i = -4\pi \lambda T\Tb~. }
A remarkable consequence \refs{\SmirnovLQW,\CavagliaODA} is that the exact $\lambda$ dependence of the energy spectrum can be written down explicitly.   Taking the theory to live on a spatial circle of circumference $L$ and using dimensional analysis to write energy eigenvalues as $E_n ={1\over L} \Ec_n(\lambda/L^2)$  one can establish the differential equation
\eqn\ab{  {1\over \pi^2} {\cal E}'_n - {\cal E}_n^2 -{2\lambda \over L^2} {\cal E}_n{\cal E}_n' +p^2 L^2=0~, }
where $p$ is the momentum of the state.
The solution yields
\eqn\ac{ E_n   =-{L\over 2\pi^2 \lambda}\left(\sqrt{1-{4\pi^2 \lambda \over L}E_{n,0}+\left({2\pi^2 \lambda \over L}p\right)^2}-1 \right)~,}
where $E_{n,0}$ is the energy of the state in the undeformed theory.

One of the main observations of \McGoughLOL\ was that the formula \ac\ arises in pure AdS$_3$ gravity by considering the quasilocal energy \refs{\BrownBR,\BrownGS} defined on a surface at finite radial location $r$. The expression appearing under the square root above indeed exhibits a marked similarity to the function appearing in the standard form of the BTZ solution.  In our conventions, the quasilocal energy is given as $E= {1\over 2\pi} \int\! d\phi \sqrt{g_{\phi\phi}} u^i u^j T_{ij}$, where $u^i$ is the timelike unit normal to the integration surface, and $T_{ij}$ is the usual boundary stress tensor \refs{\BrownBR,\BalasubramanianRE},
\eqn\ad{T_{ij} = {1\over 4 G} (K_{ij}- K h_{ij} + {1\over \ell} g_{ij} )~.}
Here $g_{ij}$ is the boundary metric, $K_{ij}$ is the extrinsic curvature, and $\ell$ is the AdS scale.  Evaluated in BTZ on a surface of fixed $r$, the quasilocal energy turns out to match \ac\ under the identification $\lambda = {4G\ell \over \pi}$; $L=2\pi r$. Note that our conventions differ from those of \McGoughLOL\ which instead give $\lambda = {4G\ell \over \pi r^2}$ and $L=2\pi$, though both result in the same dimensionless ratio ${\lambda \over L^2}$.  This dimensionless ratio is the only physical measure of the `distance' that the boundary has been moved into the bulk.   We henceforth set $\ell=1$.

As for propagation speeds, if one considers a QFT state in the deformed theory with constant $\langle T_{++}\rangle $ and $\langle T_{--}\rangle$, then small perturbations of the stress tensor can be shown to propagate at speeds
\eqn\ae{ v_{+} = 1+ 2\pi  \lambda \langle T_{++}\rangle +O(\lambda^2)~,\quad  v_{-} = 1+ 2\pi  \lambda \langle T_{--}\rangle+O(\lambda^2). }
The same propagation speeds arise in pure AdS$_3$ gravity by considering perturbations that preserve Dirichlet boundary conditions on the cutoff surface \refs{\BrattanMY,\MarolfDR}.

The following observations are useful to understand the origin and generality of these correspondences.  First of all, if we use coordinates such that $ds^2 = d\rho^2 + g_{ij}(\rho,x) dx^i dx^j$ with a cutoff surface at fixed $\rho$, then the $\rho\rho$ component of the Einstein equations is
\eqn\af{ -{1\over 2}R^{(2)} +{1\over 2} \big[ K^2- K^{ij}K_{ij}\big] - 1=0~.}
We then note that this equation applied to the stress tensor \ad\ is easily seen to imply the key trace relation \aa\ under the identification $\lambda =4G/\pi$, assuming a flat boundary metric.  This observation suffices to explain the agreement of the propagation speeds, since these can be obtained by studying linearized perturbations of the conservation equation $\nabla^i T_{ij}=0$ combined with the trace relation.   The agreement of the energy spectrum also follows readily from the Einstein equations; this time we note the the $tt$ component of the Einstein equations becomes the flow equation \ab, and hence the same solution \ac\ obtains.

To further explore the proposed correspondence, we consider the computation of stress tensor correlation functions on the two sides, focussing on two and three point functions. Elementary considerations on the QFT side yield results for the two-point functions to order $\lambda^2$, and we find, for example, $\langle T_{zz}(x)T_{zz}(0)\rangle= {c/2 \over z^4} +{5\pi^2 \lambda^2 c^2 \over 6}{1\over z^6 \zb^2}+ O(\lambda^3)$. On the bulk side, we adopt the standard AdS methodology of relating boundary stress tensor correlators to the variation of the on-shell action with respect to the boundary metric.   Perhaps surprisingly, this leads to the result that two-point functions are exactly the same whether the boundary is at a finite radial location or off at infinity as usual\foot{The commutator part of this result follows from the symplectic structure computations of \AndradeFNA.}; in particular we have $\langle T_{zz}(x)T_{zz}(0)\rangle= {c/2 \over z^4}$ where $c$ is the usual Brown-Henneaux central charge.  Does this conflict with the presence of an order $\lambda^2$ correction on the QFT side?  Under the correspondence, $\lambda =4G/\pi \sim 1/c$, and so we see that the $\lambda^2 c^2$ contribution is down by a factor of $1/c$ compared to the leading term.  On the bulk side, this corresponds to a suppression by a factor of $G$, which implies that it is a one-loop effect and therefore not seen by our classical analysis.  So our results are not in conflict with the proposed correspondence provided one compares results order by order in $1/c$, recalling that $\lambda \sim 1/c$.

To test this further, and in particular to check agreement between quantities that do receive corrections in $\lambda$, we next turn to three-point functions.   Consider the representative examples $\langle T_{z\zb}T_{zz}T_{\zb\zb}\rangle$ and $\langle T_{zz}T_{\zb\zb}T_{\zb\zb}\rangle$. These both vanish in the undeformed CFT, but get contributions of order $\lambda c^2$ in the deformed theory, and the explicit results are easily computed in conformal perturbation theory.   Since $\lambda c^2 \sim c \sim 1/G$ we expect these results to agree with a classical bulk computation, and we indeed establish precise agreement.   We similarly establish agreement for all stress-tensor three-point functions at this order.

As in the case of the two-point functions, the three-point functions in the QFT also receive higher order corrections in $1/c$, and the prediction is that these should match the corresponding loop diagrams in the bulk.  It would of course be interesting to verify the one-loop (and higher) agreement, but we leave this to future work.

On the QFT side, the trace relation \aa\  is an exact operator statement for any deformed CFT, and the energy spectrum \ac\ is similarly an exact relation governing the change in energy of all states in the original CFT.  On the other hand, the statements made on the bulk side so far apply only to pure gravity solutions. But what happens when there are nontrivial matter fields in the bulk? One can first ask whether the duality can persist as before, with the boundary QFT still being just a $T\Tb$ deformed CFT. It is easy to see that this does not work.     In terms of our previous discussion, the new issue is that the Einstein equations now pick up an extra matter stress tensor term.  So instead of getting \aa\ in the bulk we get $T^i_i = -4\pi T\Tb - t^\rho_\rho$, where $t_{ij}$ is the matter stress tensor. There is no reason for $t^\rho_\rho$ to vanish, so there is a conflict. Similarly, the energy flow equation \ab\ now gets a contribution from $t^t_t$. The quasilocal energy as a function of radial location can still be worked out explicitly in the case of a static solution, but the result is a much more complicated dependence on the radial coordinate, and the simple relation between $\lambda$ and $r$ is lost.

To gain more insight we consider scalar two-point functions.   In the bulk we compute the two-point function for a free scalar field with Dirichlet boundary conditions on the cutoff surface.   At long distance this goes over to the usual AdS correlator, but there is an infinite series of corrections to this result.   To reproduce these we need to add to the QFT action a series of double trace operators built out of the operator $O$ dual to the bulk scalar.  The presence of such additional terms in the action is consistent with our statement above regarding the change in the trace relation and energy spectrum.   We also study the effect of the leading order $T\Tb$ perturbation on the two-point function in the QFT, which turns out to yield both power law and logarithmic corrections to the correlator.  These corrections, perhaps supplemented by other interactions that involving both the stress tensor and scalar operator,  should correspond to one-loop graviton corrections in the bulk, by the same logic as in the stress tensor correlators.

To summarize our findings, it appears to us that the duality proposed in \McGoughLOL\ can successfully relate stress tensor correlators in the $T\Tb$ deformed CFT to the corresponding correlators computed in pure gravity with boundary conditions at a finite location in the bulk, although this statement remains to be checked at loop level.  On the other hand, the situation is more complicated once bulk matter is introduced.   Interactions above and beyond those of $T\Tb$ need to be introduced, and essentially fixed by hand to reproduce bulk results.

\newsec{$T\Tb$ review}

We begin by reviewing salient features of $T\Tb$ deformed conformal field theories \refs{\ZamolodchikovCE,\SmirnovLQW,\CavagliaODA}.

We define the stress tensor  via the metric variation of the Euclidean action,
\eqn\baa{ \delta S = {1\over 4\pi} \int\! d^2x \sqrt{g} T^{ij}\delta g_{ij}~.}
Given a general 2D QFT, we can define the bilocal operator
\eqn\ba{ T\Tb(x,y) = {1\over 8} T^{ij}(x)T_{ij}(y) - {1\over 8}T^i_i(x) T^j_j(y)~. }
On the flat metric $ds^2 = dzd\zb$, to which we now restrict unless stated otherwise,   this reduces to
\eqn\ba{ T\Tb(x,y) = T_{zz}(x) T_{\zb\zb}(y) - T_{z\zb}(x)T_{z\zb}(y) ~. }
As shown in \ZamolodchikovCE, this operator exhibits a remarkable OPE structure as $x\rt y$,
\eqn\bb{ T\Tb(x,y) =  O(y) + \sum_\alpha A_\alpha(x-y) \nabla_y O_\alpha(y)~.}
The (possibly divergent) functions $A_\alpha(x-y)$ multiply $y$-derivatives of local operators.  We can use this relation to identify the local operator $T\Tb(y)$ as  $O(y)$ modulo derivatives of other local operators.  Another way to say this is that
\eqn\bc{ \int\! d^2x \sqrt{g} T\Tb(x)  \equiv \lim_{\eps\rt 0} \int\! d^2x \sqrt{g} T\Tb(x,x+\eps)}
provides an unambiguous and UV finite definition of the integrated operator $\int\! d^2x \sqrt{g} T\Tb(x)$, and we adopt this definition henceforth.

Starting from a generic QFT with Euclidean action $S_0$, the $T\Tb$ deformed action is defined via the equation\foot{Note that our $\lambda$ is related to $\mu$ in \McGoughLOL\ by $\mu =4\pi^2 \lambda$, since our stress tensor differs by a factor of $2\pi$.}
\eqn\bd{ {dS(\lambda) \over d\lambda} =  \int\! d^2x \sqrt{g} T\Tb(x) }
subject to the boundary condition $S(0)=S_0$.   Importantly, the $T\Tb $ operator appearing on the right hand side is defined in terms of the stress tensor corresponding to the action $S(\lambda)$.  Hence the equation \bd\ implies a nonlinear $\lambda$-dependence for $S(\lambda)$.   We can imagine solving \bd\ by starting with a given $S(\lambda)$, computing the stress tensor of that theory, and then using $\bd$ to obtain $S(\lambda+\delta \lambda)$.

In general, for a theory with a single mass scale $\mu$ dimensional analysis yields
\eqn\bde{ \mu {dS\over d\mu} = {1\over 2\pi}\int\! d^2x \sqrt{g} T^i_i~.}
A CFT deformed by $T\Tb$ has the single scale\foot{This statement is not entirely innocuous; one should consider the possibility of scales arising through renormalization and from the possible presence of a UV cutoff, but we ignore these issues here.} $\lambda = 1/\mu^2$,  and so the relation \bd\ yields
\eqn\bdf{ T^i_i = -4\pi \lambda T\Tb~. }
Strictly speaking this result holds only under the integral since the right hand side is only defined up to total derivatives. However, \bdf\ is correct as written to first order in $\lambda$ since the operator product defining $T\Tb$ is nonsingular when the stress tensor is that of a CFT.

\subsec{Deformed free scalar action, and Nambu-Goto}

It is instructive to carry out this procedure at the classical level starting from  the action for free scalar fields (this was done in \CavagliaODA)
\eqn\bez{ S_0 = {1\over 4\pi}\sum_{n=1}^N \int\! d^2x \sqrt{g} \p^i \phi_n \p_i \phi_n~. }
First consider a single scalar field, and write the ansatz
\eqn\bfz{ S(\lambda) = \int\! d^2x \sqrt{g} \lambda^{-1}F( \lambda \p^i \phi \p_i \phi )~.}
The defining equation \bd\ becomes a differential equation for $F(z)$ which is readily solved as
\eqn\bg{ S(\lambda) = {1\over 2\pi}  \int\! d^2x \sqrt{g}\left( {1-\sqrt{1-\pi \lambda \p^i \phi \p_i \phi } \over \pi \lambda}\right) ~.}
The case of multiple scalar fields requires a more general ansatz,
\eqn\bh{ F=F(\lambda \p^i \phi_n \p_i \phi_n,\lambda^2 \p^i \phi_m \p^j \phi_m \p_i \phi_n \p_j \phi_n)~.}
This leads to a partial differential equation for $F$ which turns out to have a solution corresponding to the action
\eqn\bi{ S(\lambda) = {1\over 2\pi} \int\! d^2x { \sqrt{g}-\sqrt{\det(g_{ij}-\pi \lambda  \p_i \phi_n \p_j \phi_n) } \over \pi \lambda} ~. }
The relation \bdf\ is readily verified.
Up to an additive constant, the action \bi\ is recognized as the Nambu-Goto action written in static gauge, as is made manifest by writing
\eqn\bj{  X^0=x^0~,\quad X^n= \sqrt{-\pi \lambda} \phi_n~,\quad X^{N+1}=x^1~,}
so that
\eqn\bk{S(\lambda) = {1\over 2\pi^2 \lambda} \int\! d^2x \sqrt{\det \p_i X^A \p_j X^A}+{\rm constant}~,\quad A=0,1, \ldots N+1~.}
The Nambu-Goto action exhibits  manifest SO$(N+2)$ global symmetry, along with reparametrization invariance.   The SO$(N+2)$ symmmetry is nonlinearly realized in the gauge fixed form \bi\ due to the need to perform a compensating reparametrization to maintain static gauge.   There is no obvious {\it a priori} connection between the $T\Tb$ deformation and the existence of this global symmetry.   Of course, our discussion of the free boson theory has been purely classical, and at the quantum level one encounters the usual issues regarding the quantization of the Nambu-Goto action outside the critical dimension.  This discussion is most naturally phrased in the language of effective strings (see \AharonyIPA\ for a very clear review of the relevant issues), in which a series of higer derivative terms are added to $S(\lambda)$.   Based on physical considerations, namely that effective strings appear as solutions of Lorentz invariant theories -- e.g. as QCD strings or Nielsen-Oleson vortices -- one expects that there exists a quantization of \bk\ that preserves the SO$(N+2)$ symmetry.

Returning to the single scalar theory \bg, now in Lorentzian signature, the Hamiltonian is
\eqn\bl{ H(\lambda) = -{1\over 2\pi^2 \lambda} \int\! dx^1 \Big[\sqrt{(1-4\pi^3 \lambda \pi_\phi^2)(1-\pi \lambda (\phi')^2)}-1\Big]~.}
This illustrates that the choice of sign for $\lambda$ is quite significant;  taking $\lambda >0$ implies a rather unusual constraint on the phase space in order to preserve reality conditions.     From this perspective, $\lambda<0$ appears rather more conventional  than $\lambda>0$.

In terms of the energy and momentum of the undeformed theory,
\eqn\bm{ E_0 = \int\! dx^1 \big( \pi \pi_\phi^2 + {1\over 4\pi} (\phi')^2 \big)~,\quad p = - \int \! dx^1 \pi_\phi \phi' }
a configuration of constant $\pi_\phi$ and $\phi'$ on a circle of length $L$ has energy
\eqn\bn{H(\lambda)=-{L\over 2\pi^2 \lambda}\left(\sqrt{1-{4\pi^2 \lambda \over L}E_0+\left({2\pi^2 \lambda \over L}p\right)^2}-1 \right)~.}
This illustrates that for $\lambda>0$ configurations of sufficiently large $E_0$ for a given $p$ render the energy complex in the deformed theory.  The expression \bn\ is a classical version of the general quantum result, which we now review.

\subsec{Energy spectrum}

We start with a CFT on a spatial circle of size $L$, and assume the theory has a discrete spectrum.    Assuming $\lambda$ is the only scale present in the $T\Tb$ deformed theory, the energy of the nth state can be written
\eqn\ca{ E_n = {\Ec_n(\lambda/L^2)\over L}~.}
The momentum $p$ is integer quantized in units of $2\pi/L$ and so does not change with $\lambda$.    As shown in \refs{\SmirnovLQW,\CavagliaODA} the following differential equation holds
\eqn\cb{  {1\over \pi^2} {\cal E}'_n - {\cal E}_n^2 -{2\lambda \over L^2} {\cal E}_n{\cal E}_n' +p^2 L^2=0~. }
The solution yields
\eqn\cc{ E_n   =-{L\over 2\pi^2 \lambda}\left(\sqrt{1-{4\pi^2 \lambda \over L}E_{n,0}+\left({2\pi^2 \lambda \over L}p\right)^2}-1 \right)~,}
where $E_{n,0} = \Ec(0)/L$ is the energy of the state in the undeformed theory.  This agrees with the previous classical result \bn.  We emphasize that the assumptions going into the result \cc\ are quite minimal, essentially just that the $T\Tb$ deformed CFT exists as a theory with a single scale $\lambda$.

\newsec{AdS$_3$ gravity with a radial cutoff}

We now turn to the gravity side of the correspondence.  Most of the following section is a rederivation of results in \McGoughLOL\ from a slightly different point of view.

\subsec{Basic formulas}

The action for pure gravity in AdS$_3$ is
\eqn\da{ S = -{1\over 16\pi G} \int_M\!d^3x \sqrt{g}(R+2\ell^{-2} ) -{1\over 8\pi G}\int_{\p M}\! d^2x \sqrt{h} (  K-\ell^{-1})~.}
We work in Euclidean signature, and henceforth set the AdS radius to $1$:  $\ell=1$.  Our curvature conventions are that $R({\rm AdS}_3)=-6$.  $h_{ij}$ is the metric on the boundary.   In a coordinate system such that the metric takes the form

\eqn\db{ ds^2 = d\rho^2 + g_{ij}(x,\rho)dx^i dx^j~, }
the extrinsic curvature is
\eqn\dc{ K_{ij}={1\over 2} \p_\rho g_{ij}~. }
It is also useful to note that after integration by parts the action takes the form
\eqn\dca{ S= -{1\over 16\pi G}\int\! d^3x  \sqrt{g} \left(R^{(2)} + K^2 - K^{ij}K_{ij} +2\right)  +{1\over 8\pi G}\int_{\p M}\! d^2x \sqrt{h}~. }
Einstein's equations $R_{\mu\nu}-{1\over 2}Rg_{\mu\nu}-g_{\mu\nu}=0$ in the coordinate system \db\ take the form
\eqn\dd{\eqalign{ E^i_j&= -\p_\rho( K^i_j-\delta^i_j K) - KK^i_j+{1\over 2}\delta^i_j \big[ K^{mn}K_{mn}+K^2 \big] - \delta^i_j=0 \cr
E^\rho_j&= \nabla^i (K_{ij}-K g_{ij}) =0 \cr
E^\rho_\rho&= -{1\over 2}R^{(2)} +{1\over 2} \big[ K^2- K^{ij}K_{ij}\big] - 1=0~. }}
The on-shell variation of the action, $\delta S = {1\over 4\pi}\int\!d^2x \sqrt{h} T^{ij}\delta h_{ij}$ yields the stress tensor
\eqn\de{T_{ij} = {1\over 4 G} (K_{ij}- K g_{ij} + g_{ij} )}
which obeys $\nabla^i T_{ij} =0$ by virtue of the field equation $E^\rho_i=0$.

\subsec{Trace relation}

Recall that on the CFT side the trace relation $T^i_i =-4\pi\lambda T\Tb$ is equivalent to \bd\ which fixes the form of the deformed action.  So we would like to see this relation appearing on the gravity side as well.     Using the definition of the boundary stress tensor \de, together with the constraint equation $E^\rho_\rho=0$ we compute
\eqn\df{\eqalign{ T^i_i &= {1\over 4G}(2-K) \cr
 T\Tb &={1\over 8} (  T^{ij}T_{ij} - (T^i_i)^2) = -{1\over 64 G^2}(2-K) -{R^{(2)}\over 128 G^2}~.   }}
 This implies that on a flat boundary metric we have
\eqn\dg{ T^i_i = -16G T\Tb~.  }
Comparing to $T^i_i = -4\pi \lambda T\Tb$ we read off
\eqn\dh{ \lambda = {4G\over \pi}~.}
 We emphase that \dg\ holds for any solution of the Einstein equations with a flat boundary metric.  In \McGoughLOL\ the relation between the deformation parameter $\lambda$ and bulk quantities involves the radial location $r$ of the cutoff surface, whereas the relation \dh\ involves no such thing.  $T^i_i$ and $T\Tb$ are both coordinate independent objects, so the relation between them cannot involve an arbitrary radial coordinate. However, a radially dependent expression for $\lambda$ will emerge naturally below when we consider the spatial circle to have a specified size $L$.

 \subsec{Propagation speed}

 The fact that under the dictionary \dh\ we get the same trace relation in CFT and gravity immediately implies that we will get agreement for the propagation speed of stress energy perturbations.  This follows because the propagation speed is derived using just the conservation equations and the trace relation.   Namely, on a flat metric $ds^2 =dzd\zb$ these equations are
 \eqn\di{\eqalign{ &\p_{\zb}T_{zz} +\p_z T_{z\zb} =0 \cr
 & \p_{z}T_{\zb\zb} +\p_{\zb} T_{z\zb} =0\cr
  & T_{z\zb} +\pi \lambda \big( T_{zz}T_{\zb\zb} -(T_{z\zb})^2 \big)=0~.}}
Upon linearizing these equations (after converting to Lorentzian signature) around a background of constant $\langle T_{ij}\rangle$ it is straightfoward to show that perturbations propagate at speeds
\eqn\dj{ v_{+} = 1+ 2\pi  \lambda \langle T_{++}\rangle +O(\lambda^2)~,\quad  v_{-} = 1+ 2\pi  \lambda \langle T_{--}\rangle+O(\lambda^2) }
in agreement with results stated in \McGoughLOL.  The superluminal nature of these speeds for $\lambda >0$ has been discussed in \refs{\BrattanMY,\MarolfDR}. In the bulk this can be understood simply as coming from the coordinate transformation needed to put the metric on the constant $r$ surface in standard form.

\subsec{Energy spectrum}

We now consider the Euclidean BTZ metric
\eqn\dk{ds^2 = {dr^2 \over f(r)^2} +f(r)^2 dt^2 + r^2(d\phi -i\omega(r)dt)^2 }
with
\eqn\dl{ f(r)^2 = r^2-8GM +{16G^2J^2 \over r^2}~,\quad \omega(r) = {4GJ\over r^2}~.}
It is convenient here to focus on the dimensionless ``proper energy"
\eqn\dm{\Ec  = EL = {L\over 2\pi} \int\! d\phi \sqrt{g_{\phi\phi}} u^i u^j T_{ij}~,}
where $u^i$ is the unit normal to a constant $t$ slice of the boundary,
\eqn\dn{ u^t = {1\over f}~,\quad u^\phi = {i\omega \over f}~.}
$L=\int\! d\phi \sqrt{g_{\phi\phi}}$ is the proper size of the spatial circle on the boundary.  We have $L=2\pi r$ and $\lambda = {4G\over \pi}$.
We now compute
\eqn\do{\eqalign{\Ec & = {L^2 \over 2\pi^2 \lambda } \big(1-r^{-1}f(r)\big)\cr
& =  {L^2 \over 2\pi^2 \lambda }    \left(1-\sqrt{1-{4\pi^2 \lambda \over L}M+\left({2\pi^2 \lambda \over L}J\right)^2}~\right)~,}}
which agrees with \cc\ under the identification $M=E_0$, $J=p$.

Another way to arrive at this conclusion is to observe that the flow equation \cb\ follows from Einstein's equations.   For simplicity, consider the case $p=0$ corresponding to a static metric.  Writing a general static metric in the form   $ds^2 = dr^2/f(r)^2 + g(r)^2dt^2 + r^2 d\phi^2$, and without using the Einsten equations we compute $\Ec  = {L^2 \over 2\pi^2 \lambda }    \big(1-r^{-1}f(r)\big)$.  The Einstein equation $E^t_t=0$ is then easily seen to be nothing but \cb.

\newsec{Correlation functions in the deformed CFT}

A CFT deformed by $T\Tb$ has an action that obeys ${dS \over d\lambda } = \int\! d^2x \sqrt{g} T\Tb$, which implies $S(\lambda) = S_{0} + \lambda  \int\! d^2x \sqrt{g} [T\Tb]_0 + O(\lambda^2)$, where $S_0$ and $[T\Tb]_0$ are the action and perturbing operator of the undeformed CFT.    We now ask whether it is sensible to compute correlation functions in the deformed theory.  The obvious issue is that since we are perturbing the CFT by an irrelevant (in the RG sense) operator, we potentially have to deal with all the issues associated with non-renormalizable theories.  In particular, we could take the effective field theory point of view, imposing a UV cutoff and computing correlators in the presence of an infinite number of counterterms, each with an arbitrary coefficient.   However, if we restrict attention to correlation functions of the stress tensor the situation is much more favorable and it is possible to draw some universal conclusions.

We confine our analysis to perturbation theory in $\lambda$; the definition of correlators at the non-perturbative level is of course a much more difficult question. The first point to recall is that the operator $\int\! d^2x \sqrt{g} T\Tb$ defined as in \bc\ is finite and unambiguous, and so no dependence on a renormalization scale enters. Next, let's recall the standard statement that conserved currents, such as the stress tensor, are not renormalized in perturbation theory.  The usual argument for this (e.g. chapter 4 of \coleman) involves deriving a Ward identity by making an infinitesimal symmetry transformation in the renormalized path integral.   The Ward identity takes the schematic form
\eqn\dt{ \langle \p_\mu J^\mu(y) \phi(x_1) \ldots \phi(x_n) \rangle = \delta(y-x_1)\langle  \delta \phi(x_1) \ldots \phi(x_n) \rangle + \ldots + \delta(y-x_n) \langle   \phi(x_1) \ldots \delta \phi(x_n) \rangle~,}
where $\delta \phi$ is the transformation generated by the current. Since the correlators on the right hand side are those of renormalized fields they are by definition finite, and hence so too is the left hand side.  Thus, the current $J^\mu$ obtained by applying Noether's theorem to the renormalized action has the property that $\p_\mu J^\mu$ has finite correlators with renormalized fields.   Thus, up to the possible addition of an identically conserved vector operator, the same is true of $J^\mu$.  We should note that such identically conserved vector operators can indeed make an appearance; for example they do so for the U(1) current in QED, where the operator is $V^\mu = \p_\nu F^{\mu\nu}$ \CollinsNJ.

Here we are concerned with the stress tensor, corresponding to $J^\mu = T^{\mu\nu}\xi_\nu$, where $\xi_\nu$ is a Killing vector. We will assume that there exists no identically conserved tensor that can mix under renormalization with the stress tensor; this should hold generically, since any such operator would need to have scaling dimension equal to precisely $2$.  Under this assumption, the stress tensor defined in the usual way from the renormalized action will have finite correlators with renormalized fields.  In particular, combining this with the statement about the finiteness of the deforming operator, we conclude that stress tensor correlators are finite and independent of renormalization scale (they do of course depend on the dimensionful scale $\lambda$).

We should note that this argument assumes the existence of a stress tensor obeying the Ward identity.  However, a complete argument should give a prescription for defining this object.  We would like to define the stress tensor as the variation of the action with respect to the background metric, but the issue here is that the $T\Tb$ perturbation is only unambiguously defined on a flat background.  Therefore, we leave to the future a definitive answer to the question of whether all stress tensor correlators can be computed perturbatively in $\lambda$ without ambiguity. Here we will only consider low point correlators at the first nontrivial order in perturbation theory, where this subtlety does not appear to arise.

\subsec{Two-point functions}

We first place general constraints on the form of the two-point functions; these hold equally well in CFT and in the bulk.
On the metric $ds^2=dzd\zb$ and in the presence of the $T\Tb$ deformation, dimensional analysis, and translation/rotational symmetry imply
\eqn\dod{\eqalign{ \langle T_{zz}(x)T_{zz}(0)\rangle_\lambda & = {1\over z^4} f_1(y) \cr
\langle T_{zz}(x)T_{z\zb}(0) \rangle_\lambda& = {1\over z^3 \zb} f_2(y) \cr
\langle T_{zz}(x)T_{\zb\zb}(0) \rangle_\lambda& = {1\over z^2 \zb^2} f_3(y) \cr
\langle T_{z\zb}(x)T_{z\zb}(0)\rangle_\lambda & = {1\over z^2 \zb^2} f_4(y) }}
and then also by symmetry we have
\eqn\doe{\eqalign{ \langle T_{\zb\zb}(x)T_{\zb\zb}(0)\rangle_\lambda & = {1\over \zb^4} f_1(y) \cr
\langle T_{\zb\zb}(x)T_{z\zb}(0)\rangle_\lambda & = {1\over z \zb^3} f_2(y)~.  }}
Here the dimensionless variable $y$ is
\eqn\doea{  y = {z\zb \over \lambda}~.}
Demanding stress tensor conservation implies
\eqn\doh{\eqalign{f_1'+y^3 \left(f_2 \over y^3\right)'&=0\cr \left(f_2 \over y\right)'+ y \left(f_3 \over y^2\right)'&=0 \cr
 \left(f_2 \over y\right)'+ y \left(f_4 \over y^2\right)'&=0~. }}
Since we have a CFT perturbed by an irrelevant deformation, correlators should go over to their CFT values at long distance, which implies that we are looking for solutions with boundary conditions
\eqn\dol{ f_1 \rt {c\over 2}~,\quad {1\over y^2} f_2  \rt 0~,\quad {1\over y^2} f_3  \rt 0~,\quad {1\over y^2} f_4  \rt 0~,\quad y \rt \infty }

\subsec{Two-point functions in deformed CFT}

It is now simple to work out the deformed CFT two-point function to order $\lambda^2$. We consider operators at distinct points corresponding to ignoring possible contact terms.   Recall that we have the operator equation $T_{z\zb}=-\pi \lambda T\Tb $, and so
\eqn\dom{ T_{z\zb}=-\pi \lambda  T_{zz}T_{\zb\zb} + O(\lambda^2) }
where the stress tensors appearing on the right hand side are those of the undeformed CFT.
So we can compute $\langle T_{z\zb}(x) T_{z\zb}(0)\rangle$ to order $\lambda^2$ by using the above relation and evaluating correlators in the undeformed theory.   This gives
\eqn\don{ f_4(y)  = {\pi^2 c^2 \over 4 y^2} + \ldots }
where $\ldots$ are terms with further $1/y$ suppression.  The conservation equations \doh\  and boundary conditions \dol\ now fix the leading behavior of the other functions, and the result is
\eqn\dona{\eqalign{\langle T_{zz}(x)T_{zz}(0)\rangle_\lambda & =  {c\over 2 z^4} +{5\pi^2 \lambda^2 c^2\over 6} {1\over z^6 \zb^2} + \ldots \cr
 \langle T_{zz}(x)T_{z\zb}(0)\rangle_\lambda & = -{\pi^2 \lambda^2 c^2\over 3}{1\over z^5 \zb^3}+\ldots  \cr
 \langle T_{zz}(x)T_{\zb\zb}(0)\rangle_\lambda & =  {\pi^2 \lambda^2 c^2\over 4} {1\over z^4\zb^4} +\ldots  \cr
 \langle T_{z\zb}(x)T_{z\zb}(0)\rangle_\lambda & =   {\pi^2 \lambda^2 c^2 \over 4}{1\over z^4\zb^4} +\ldots   } }
Higher order corrections can be worked out in conformal perturbation theory.

Note that there are no corrections at order $\lambda$.  The correlator $\langle T_{zz}(x) T_{\zb\zb}(0)\rangle$ would seem to get an order $\lambda$ contribution by bringing down one $\lambda T\Tb$ interaction vertex, but the corresponding integral turns out to vanish.

\subsec{Three-point functions in deformed CFT}

We start by considering  a couple of examples that can be easily computed.   The simplest nontrivial three-point function result is the order $\lambda$ contribution to the correlator  $\langle T_{z\zb}(x_1) T_{zz}(x_2)T_{\zb\zb}(x_3)\rangle$.  We simply use the relation \dom\ to obtain
\eqn\donb{ \langle T_{z\zb}(x_1) T_{zz}(x_2)T_{\zb\zb}(x_3)\rangle_\lambda = -{\pi \lambda c^2\over 4} {1\over (z_1-z_2)^4 (\zb_1-\zb_3)^4} + O(\lambda^2)~.}

A slightly less trivial example is provided by the order $\lambda$ contribution to the correlator $\langle T_{zz}(x_1) T_{\zb\zb}(x_2) T_{\zb\zb}(x_3)\rangle$.  Using $\sqrt{g} d^2x = {1\over 2}d^2z$, $\p_{\zb} {1\over z} = 2\pi \delta^2(z)$,\foot{We should note that the justification of the standard rule $\p_{\zb} {1\over z} = 2\pi \delta^2(z)$ is not totally straightforward in this context.  It is based on cutting out a disk of radius $\eps$ around the singularity, computing, and then taking $\eps \rt 0$. But recall that the unintegrated $T\Tb$ operator is defined only up to potentially divergent derivative terms, and we should thus worry about such contributions at the boundary of the disk.  A complete treatment of perturbation theory would need to confront this.      }  and repeatedly integrating by parts,  we have
\eqn\drl{  \eqalign{ & \langle T_{zz}(x_1) T_{\zb\zb}(x_2) T_{\zb\zb}(x_3)  \rangle_\lambda \cr
&  \quad  =-{\lambda}   \int\! d^2x \sqrt{g}  \langle T_{zz}(x_1) T_{\zb\zb}(x_2) T_{\zb\zb}(x_3) T_{zz}(x)T_{\zb\zb}(x)\rangle \cr & \quad  =
-{\lambda c^2 \over 4} \int\! d^2z  {1\over (z-z_1)^4 (\zb-\zb_2)^2(\zb-\zb_3)^2(\zb_2-\zb_3)^2}  \cr
&   \quad = {\lambda c^2 \over 12} \int\! d^2z \p_z  {1\over (z-z_1)^3}{1\over  (\zb-\zb_2)^2(\zb-\zb_3)^2(\zb_2-\zb_3)^2} \cr
&  \quad = - {\lambda c^2 \over 12} \int\! d^2z  {1\over (z-z_1)^3}\p_z {1\over (\zb-\zb_2)^2(\zb-\zb_3)^2(\zb_2-\zb_3)^2}  \cr
 &  \quad =   {\pi \lambda c^2 \over 6}\int\! d^2z  {1\over (z-z_1)^3} \p_{\zb}\delta^2(z-z_2) {1\over (\zb-\zb_3)^2(\zb_2-\zb_3)^2} +  ( x_2 \leftrightarrow  x_3)\cr
 & \quad = {\pi \lambda c^2 \over 3}\int\! d^2z  {1\over (z-z_1)^3} \delta^2(z-z_2) {1\over  (\zb-\zb_3)^3(\zb_2-\zb_3)^2} +  ( x_2 \leftrightarrow  x_3)\cr
 & \quad  = - {\pi \lambda c^2 \over 3}  {1\over (z_1-z_2)^3} {1\over (\zb_2-\zb_3)^5} + ( x_2 \leftrightarrow  x_3) }}

A little thought reveals that all three-point functions at order $\lambda$ are fixed by simple considerations.  Consider a correlator involving $T_{z\zb}$.  We can evaluate this to order $\lambda $ by using \dom\ and the undeformed correlators.   Noting the symmetry under $z\leftrightarrow \zb$, this just leaves the $\langle T_{zz}T_{\zb\zb}T_{\zb\zb}\rangle$, which we worked out in \drl, and $\langle T_{zz}T_{zz}T_{zz}\rangle$.  But the latter correlator clearly has no order $\lambda$ contribution, since the $T_{\zb\zb}$ operator in the interaction term has nothing to contract against.

\newsec{Correlators in cutoff AdS}

We now turn to the computation of stress tensor correlation functions in the bulk.  We assume the same basic framework as in standard AdS correlator computations.  Namely, we compute the on-shell bulk action as a functional of the metric on the boundary, and then obtain correlators by taking functional derivatives,
\eqn\ra{ \langle T_{i_1j_1}(x_1) \ldots T_{i_nj_n}(x_n)\rangle = (4\pi)^n {\delta^n S[h] \over \delta h_{i_1j_1}(x_1) \ldots \delta h_{i_nj_n}(x_n)}~.}
This prescription has the important virtue that diffeomorphism invariance of the action implies that these correlators obey the correct conservation laws / Ward identities.

We consider metrics of the form
\eqn\rb{ ds^2 = {dy^2 +dzd\zb \over y^2} +\eps g^{(1)}_{ij}(y,z,\zb)dx^i dx^j + \eps^2 g^{(2)}_{ij}(y,z,\zb)dx^i dx^j + \ldots ~,}
and perturbatively solve the Einstein equations subject to the boundary condition
\eqn\rc{ g^{(1)}_{ij}(y=1,z,\zb) = h_{ij}(z,\zb)~.}
We are placing the boundary at $y=1$; there is no loss of generality in this choice in the sense that any fixed $y$ surface can be brought to $y=1$ by a coordinate transformation that preserves the background metric.

\subsec{Two-point function}

We read off the two-point function via
\eqn\rd{ T_{ij}(x) = {1\over 4\pi} \int\! d^2x' \sqrt{g^{(0)}(x')} \langle T_{ij}(x) T^{kl}(x') \rangle h_{kl}(x')~,}
where as usual
\eqn\re{ T_{ij} = {1\over 4G}(K_{ij}-Kg_{ij}+g_{ij})~.}
The Einstein equations are $E_{\mu\nu} = R_{\mu\nu}-{1\over 2}Rg_{\mu\nu}=g_{\mu\nu}=0$.  $E_{y\mu}$ are constraint equations, and once imposed at $y=1$ they are automatically obeyed for all $y$ by virtue of the ``dynamical" equations $E_{ij}=0$.

To compute the two-point functions we need only consider the Einstein equations to first order in $\eps$. The dynamical equations $E_{ij}=0$  are
\eqn\rf{\eqalign{  y \p_y^2 g_{ij}^{(1)}+3\p_y g_{ij}^{(1)}=0~, }}
so that we have
\eqn\rg{\eqalign{  g_{ij}^{(1)}(y,z,\zb) &=  \left( {1\over y^2}-1\right) f_{ij}^{(1)}(z,\zb)   +h_{ij}(z,\zb)~.      }}

To compute $\langle T_{zz} T_{zz}\rangle$ we set $h_{zz}=h_{z\zb}=0$.  The constraint equation $E_{yy}$ gives
\eqn\rh{ E_{yy}|_{y=1} = -4(f^{(1)}_{z\zb} -{1\over 2}\p_z^2 h_{\zb\zb}) =0\quad \Rightarrow \quad f^{(1)}_{z\zb}  = {1\over 2}\p_z^2 h_{\zb\zb}~.}
The remaining constraint equations are
\eqn\ri{\eqalign{ E_{y\zb}|_{y=1}& = -y(-2\p_z f^{(1)}_{\zb\zb}+2\p_z h_{\zb\zb}+ \p_z^2 \p_{\zb} h_{\zb\zb})\eps =0 \cr
 E_{yz}|_{y=1}& =y(2\p_{\zb} f^{(1)}_{zz}-\p_z^3 h_{\zb\zb})\eps=0~,}}
yielding
\eqn\rj{\eqalign{ f^{(1)}_{\zb\zb} &= h_{\zb\zb} +{1\over 2} \p_z \p_{\zb} h_{\zb\zb} \cr
 f^{(1)}_{zz}(z) &= {3\over 2\pi} \int\! d^2z' {1\over (z-z')^4 } h_{\zb\zb}(z',\zb')~.}}
Using $T_{zz} ={1\over 4G}f^{(1)}_{zz}\eps$ we read off the correlator from \rd\ as
\eqn\rk{ \langle T_{zz}(z) T_{zz}(z')\rangle={c/2\over (z-z')^4} }
where $c=3/2G$ is the Brown-Henneaux central charge.   On the other hand, since $f^{(1)}_{z\zb}$ and $f^{(1)}_{\zb\zb}$ are both local functions of $h_{\zb\zb}$  the corresponding  correlators $\langle T_{zz}(z) T_{\zb\zb}(z')\rangle$ and $\langle T_{zz}(z) T_{z\zb}(z')\rangle$ vanish up to contact terms.

Recalling the analysis in section 4 we see that this computation is sufficient to fix all the two-point functions, and in particular we find that besides the result \rk\ and the corresponding result for $\langle T_{\zb\zb}T_{\zb\zb}\rangle$ all other two-point functions vanish up to contact terms, a result which is easily verified by repeating the previous computation for the other cases.   So at this order  the two-point functions are precisely those of a CFT and show no sign of the $\lambda$ deformation.  We note that the commutator part of this result follow from the symplectic structure computed in \AndradeFNA.

As we discussed in the introduction, this makes perfect sense when we recall that our classical analysis just gives the contribution to correlators proportional to $c$. Since $\lambda \sim 1/c$, the correction terms appearing in \dona\ are of order $c^0$, and hence correspond to one-loop effects in the bulk.    In order for a classical computation to exhibit $\lambda$ dependence, we need to turn to the three-point functions.



\subsec{Three-point functions}

We  follow the same strategy to compute  three-point functions. The intermediate algebra is a bit messy and unilluminating so we do not show all details.

We first consider $\langle T_{z\zb}(x_1) T_{zz}(x_2) T_{\zb\zb}(x_3)\rangle$.  We proceed by activating $h_{zz}$ and $h_{\zb\zb}$ and extracting the contribution to $T_{z\zb}$ proportional to the cross term $h_{zz}h_{\zb\zb}$. Using $T_{z\zb} = -{1\over 8G} \p_y g^{(2)}_{z\zb}\eps^2$ we find
\eqn\rl{ T_{z\zb}(x_1) = -{1\over 4G} f^{(1)}_{\zb\zb}(x_1)f^{(1)}_{zz}(x_1)+{\rm local}~,}
where the $f^{(1)}$ are given by the same linearized computation as appeared in the two-point function computation,
\eqn\rmz{\eqalign{f^{(1)}_{\zb\zb}(\zb_1) &= {3\over 2\pi} \int\! d^2z_2 {1\over (\zb_1-\zb_2)^4 } h_{zz}(z_2,\zb_2)\cr
 f^{(1)}_{zz}(z_1) &= {3\over 2\pi} \int\! d^2z_3 {1\over (z_1-z_3)^4 } h_{\zb\zb}(z_3,\zb_3)~.}}
Recalling that in our two-point function computation we had $T_{zz} ={1\over 4G}f^{(1)}_{zz}\eps$  (along with the corresponding result for $T_{\zb\zb}$) we see that \rl\ implies
\eqn\rn{\eqalign{  \langle T_{z\zb}(x_1) T_{zz}(x_2) T_{\zb\zb}(x_3)\rangle & = -4G \langle T_{zz}(x_1)T_{zz}(x_2)\rangle \langle \langle T_{\zb\zb}(x_1) T_{\zb\zb}(x_3)\rangle \cr
& = -{\pi \lambda c^2 \over 4} {1\over (z_1-z_2)^4(\zb_1-\zb_3)^4}~,}}
in agreement with the $O(\lambda)$ CFT result written in \donb.     This agreement is not a surprise, as it follows from the fact that our computations respects the trace condition $T^i_i =-4\pi \lambda T\Tb$, but it is a good computational check.

Next we consider $\langle T_{zz}(x_1)T_{\zb\zb}(x_2)T_{\zb\zb}(x_3)\rangle$.  We turn on $h_{zz}$ and study the response of $T_{zz}$ at second order.    We find
\eqn\ro{ \p_{\zb} T_{zz}(x_1) = -{1\over 8G} f^{(1)}_{\zb\zb}(x_1) \p_z^2 \p_{\zb} h_{zz}(x_1) +{\rm local}~,}
which we solve as
\eqn\rp{ T_{zz}(x_1) = -{3\over 4\pi^2 G} \int\! d^2 z_2 d^2 z_3 {h_{zz}(x_2) h_{zz}(x_3)\over (z_1-z_2)^3(\zb_2-\zb_3)^5}~.  }
This yields
\eqn\rq{ \langle T_{zz}(x_1)T_{\zb\zb}(x_2)T_{\zb\zb}(x_3)\rangle = -{3\over G}  {1\over (z_1-z_2)^3} {1\over (\zb_2-\zb_3)^5} + ( x_2 \leftrightarrow  x_3)~,}
which agrees with $\drl$ upon using $\lambda =4G/\pi$ and $c=3/2G$.

We can now argue that all three-point functions will match at this order.  Given any correlator involving an insertion of $T_{z\zb}$, we compute it by evaluating $\langle T_{z\zb}\rangle$ in the presence of sources for the other operators; by sources we mean variations of the boundary metric.  Since there is no source for the $T_{z\zb}$ whose value we are computing, we can use the Einstein equations for a flat metric to replace $T_{z\zb} \rt -\pi \lambda T\Tb$, and then use known results for the $\lambda^0$ correlators.  Such correlators therefore match those in the deformed CFT, since the same logic applies there.  This just leaves the correlator in \rq, which we found to match, along with $\langle T_{zz}T_{zz}T_{zz}\rangle$.  But there is no order $\lambda$ correction to this correlator at the classical level, and so we just have the undeformed CFT result, which matches the CFT at this order.

\newsec{Including matter in the bulk}

What makes the $T\Tb$ deformation on the CFT side especially interesting is its universality: it can be applied to any 2D QFT.   However, on the bulk side our discussion has so far been limited to solutions of pure gravity with a negative cosmological constant.  The obvious question is whether the appealing dictionary relating the two sides can be extended to the case where nontrivial matter fields appear on the bulk side.  We first give general arguments that the correspondence must be modified when we consider bulk solutions in the presence of classical matter fields that deform the geometry.

To start with, the Einstein equations written in \dd\  in the presence of matter are
\eqn\dpz{ E^\mu_\nu = -4G t^\mu_\nu~, }
where $t_{\mu\nu}$ is the matter stress tensor. We are supposed to solve \dpz, along with the matter field equations, subject to Dirichlet boundary conditions on a cutoff surface.
The natural Dirichlet problem is to hold the metric fixed at the boundary, here taken to be a flat metric, and to demand that matter fields are constant on the boundary.  More precisely, we demand that the matter fields on the boundary are invariant under coordinate transformations of the boundary.   These conditions ensure that the boundary stress tensor, defined exactly as before according to \de, is covariantly conserved.  This is because covariant conservation follows from diffeomorphism invariance: $\delta S = 4\pi \int\! \sqrt{g} T^{ij}\delta g_{ij}$ vanishes under $\delta_\xi g_{ij} = \nabla_{(i } \xi_{j)}$ provided the matter data on the boundary obeys $\delta_\xi \Phi=0$, and then conservation follows upon integration by parts.

As we have noted, the basic equation defining the $T\Tb$ deformation in CFT is the trace relation $T^i_i = -4\pi \lambda T\Tb$, and we saw that in the bulk this followed from the Einstein equation $E^\rho_\rho$, valid  in the absence of matter.    In the presence of matter we have $E^\rho_\rho = -4Gt^\rho_\rho$ which leads to
\eqn\dq{ T^i_i =-16GT\Tb -t^\rho_\rho~.}
In general, under our boundary conditions there is no reason why $t^\rho_\rho$ should vanish; for instance a scalar field that varies only in the radial direction will generically yield $t^\rho_\rho\neq 0$.   Therefore, we find  that imposing Dirichlet boundary conditions yields a stress tensor that does not respect the defining equation governing the $T\Tb$ deformation.

To elaborate on this, we can generalize the previous energy computation to include matter fields.  This can be done quite explicitly if we assume a static rotationally symmetric configuration for the metric and matter fields.  The bulk metric can be found explicitly in terms of the matter stress tensor; this is essentially the content of Birkhoff's theorem in this context.   Defining the radial coordinate via $g_{\phi\phi}=r^2$, we find
\eqn\dr{ \Ec = {\pi \over 2G} r^2 \left(1-\sqrt{1-{8GM(r)\over r^2}}~\right)}
where
\eqn\ds{ M(r) = E_0+ \int_r^\infty\! r' t^t_t(r') dr'~. }
It is evident that there is no simple correspondence with the CFT result \cc; in particular $\lambda$ cannot be related in any simple way to the radial coordinate $r$.
We also note that the computation of the propagation speed of perturbations will not match, since this agreement was based on the trace relation agreeing between the two sides.
Rather than pursue this avenue further, we instead turn to a discussion of free scalar fields propagating on a pure gravity background.

\newsec{Scalar correlators}

Our goal here to see what must be done on the QFT side of the duality in order to reproduce the simplest matter correlation function in the bulk, namely the two-point function of a free scalar.

\subsec{Two-point function of bulk free scalar}

The scalar two-point function is computed as usual except that we impose Dirichlet boundary conditions on a surface at finite $y=y_0$, where the metric is $ds^2 = (dy^2+dx^i dx^i)/y^2$.   The on-shell action is
\eqn\sa{\eqalign{ S&= {1\over 2} \int\! d^3 x\sqrt{g} \big( (\p \phi)^2 + m^2 \phi^2 \big) \cr
& =   {1\over 2} \int\! d^2x {1\over y_0} \phi \p_y \phi~.        }}
The wave equation has plane wave solutions
\eqn\sb{ \phi_p(y,x) = {y K_\nu(py)\over y_0 K_\nu(py_0)}\phi(p)e^{ip_i x^i} ~,\quad  \nu = 2h-1~,\quad p=\sqrt{p_ip_i}~.}
We assume a generic mass so that $\nu$ is typically not an integer.
The action is then
\eqn\sc{ S = {1\over 2}\int\! {d^2p \over (2\pi)^2} {1\over y_0} {\p_y \big( yK_\nu(py)\big)|_{y_0} \over y_0 K_\nu(py_0)} \phi(p) \phi(-p)~,}
corresponding to the two-point function
\eqn\sd{ \langle \phi(p)\phi(p')\rangle = {1\over y_0}  {\p_y \big( yK_\nu(py)\big)|_{y_0} \over y_0 K_\nu(py_0)} (2\pi)^2 \delta^{(2)}(p+p')~.}
We recall that the Bessel function has an expansion for small argument of the form
\eqn\se{ K_\nu(x) = x^{-\nu} \sum_{k=0}^\infty a_k x^{2k} + x^{\nu} \sum_{k=0}^\infty b_k x^{2k}~.}
The correlator then has the structure
\eqn\sf{ \langle \phi(p)\phi(p')\rangle =  \left[ (p^2)^\nu g_1(p^2) + (p^2)^{2\nu} g_2(p^2)+\ldots  \right] (2\pi)^2 \delta^{(2)}(p+p')~,}
where the functions $g_k(p^2)$ (which are easily worked out)  are analytic at the origin.   We suppressed the dependence on $y_0$, since we can in any case set $y_0=1$ by a coordinate transformation that preserves the metric.  The leading piece as $p\rt 0$  is $(p^2)^\nu$, which is the result in the undeformed CFT.

We can also examine the short distance behavior.  Recall that $K_\nu(x) \sim \sqrt{\pi \over 2x}e^{-x}(1+ {a\over x} + \ldots)$ as $x\rt \infty$.  This yields $\langle \phi(p) \phi(-p)\rangle \sim 1/p$, corresponding to a $1/x$ short distance behavior.  Quantum corrections to this result are expected to be important.

\subsec{Scalar two-point function in CFT with double trace perturbations}

In AdS/CFT the dual to a free scalar of mass $m$ in the bulk is a ``generalized free field": a scalar operator $O$ of dimension $2h$ set by  $m^2=4h(h-1)$ whose correlation functions factorize into products of two-point functions.  In a CFT, the two point function is $\langle O(x)O(y)\rangle \sim |x-y|^{-4h}$.

We would now like to reproduce the bulk two-point function computed on a cutoff surface in the bulk.  As noted by \FaulknerJY, this can be accomplished by adding double trace interactions to the action.  We essentially reproduce their work below.

It is convenient to work in momentum space and normalize our scalar operator in the original CFT such that
\eqn\sg{ \langle O(p) O(p')\rangle =  (p^2)^{\nu} (2\pi)^2 \delta^{(2)}(p+p')~. }
We then include  a general double trace term in the action
\eqn\sh{ S_{O^2}  = \int\! {d^2q \over  (2\pi)^2}{1\over 2} f(q^2)O(q)O(-q)~. }
The two-point function in the deformed theory is easily computed by thinking of \bb\ as an interaction and using the assumed factorization of correlators.  Summing the geometric series gives
\eqn\si{\eqalign{    \langle O(p) O(p')\rangle &=
     (p^2)^{\nu} (2\pi)^2 \delta^{(2)}(p+p') - (p^2)^{2\nu} f(p^2) (2\pi)^2 \delta^{(2)}(p+p') + \ldots \cr
& =   {    (p^2)^{\nu} \over 1+  (p^2)^{\nu}f(p^2)}  (2\pi)^2 \delta^{(2)}(p+p')~. } }
Comparing with \sf\ we can choose $f(p^2)$ to get agreement, but the needed function is necessarily non-analytic at the origin,
\eqn\sj{ f(p^2) = (p^2)^{-\nu} f_{1}(p^2) + f_2(p^2) + (p^2)^\nu f_3(p^2) + \ldots }
where the $f_k(p^2)$ are analytic.    Note that they are also of order 1; i.e., they are not suppressed by powers of $c$.  This gives another argument that such effects cannot be reproduced by the $T\bar T$ deformation alone.

The above non-analyticity implies non-locality in position space. This may be disappointing in comparison with the $T \bar T$ deformation, but is expected from e.g. the known failures of boundary-correlator micro-locality and strong subadditivty of entropy \HH\ that arise when the bulk is subjected to a strict radial cutoff.  This particular non-locality was found previously in both \HeemskerkHK\ and \FaulknerJY.

\subsec{Effect of $T\Tb$ deformation}

We now consider the first order correction to the scalar two-point function due to the $T\Tb$ deformation.  Here it is easier to  proceed in position space.  Using the CFT result, as fixed by the OPE,
\eqn\sk{ \langle O(x_1) O(x_2) T_{zz}(x)  T_{\zb\zb}(x)\rangle = h^2 \left( (x_1-x_2)^2 \over (x-x_1)^2(x-x_2)^2  \right)^2\langle O(x_1) O(x_2) \rangle~,}
the first order correction to the two-point function is
\eqn\sl{ \langle O(x_1) O(x_2) T_{zz}(z) T_{\zb\zb}(\zb)\rangle_\lambda  =  \lambda h^2 \langle O(x_1) O(x_2)  \rangle_0 (x_1-x_2)^4 \int\! { d^2x \over [(x-x_1)^2(x-x_2)^2]^2 }~.}
Introducing a Feynman parameter the integral is
\eqn\sm{ I=\int\! { d^2x \over [(x-x_1)^2(x-x_2)^2]^2 }=6 \int_0^1\! d\alpha \alpha(1-\alpha)  \int\! { d^2x \over [x^2 +(x_1-x_2)^2\alpha(1-\alpha) ]^4 } ~.}
The integral is UV divergent and so we apply dimensional regularization: $d^2x \rt d^{d}x$, yielding
\eqn\sn{\eqalign{ I& = \pi^{d/2} \Gamma(4-{d\over 2}){\Gamma({d\over 2}-2)^2\over \Gamma(d-4)} (x_1-x_2)^{d-8}\cr
& =-{16 \pi \over \eps}{1\over  (x_1-x_2)^6}  +{4\pi \over (x_1-x_2)^6} (2\ln [\pi (x_1-x_2)^2] -5+2\gamma) +O(\eps)~.   }}
where we wrote $d=2-\eps$ and expanded, and wrote $\gamma$ as the Euler constant.   So we have
\eqn\so{\eqalign{& \langle O(x_1) O(x_2) T_{zz}(z) T_{\zb\zb}(\zb)\rangle_\lambda  \cr
& \quad =  \lambda h^2  \langle O(x_1) O(x_2)  \rangle_0 \left(  -{16 \pi \over \eps}{1\over  (x_1-x_2)^2}  +{4\pi \over (x_1-x_2)^2} (2\ln [\pi \mu(x_1-x_2)^2] -5+2\gamma) \right)~.}}
The divergence indicates that $O$ mixes with $\nabla^2 O$ under renormalization, so we define a renormalized operator as $O_R = O - {A\lambda \over \eps}\nabla^2 O$, with the numerical coefficient $A$ chosen to cancel the pole.  The structure of the renormalized correlator is then
\eqn\sp{ \langle O_R(x_1) O_R(x_2)\rangle =  \langle O(x_1) O(x_2)  \rangle_0 \left( 1+ {C_1 \lambda \over (x_1-x_2)^2} +{C_2 \lambda \ln [\mu(x_1-x_2)]  \over (x_1-x_2)^2}    \right)~.}

In bulk language, this result corresponds to a one-loop correction due to a graviton loop, since $\lambda \sim G$.   We also note the appearance $\ln(x_1-x_2)$, whereas we saw above that the tree level correlator contains only power law terms for generic mass.

\newsec{Conclusion}

In this work we have further explored the cutoff AdS / $T\Tb $ deformed CFT duality.   Some of the basic features of the deformed CFT, namely the trace relation $T^i_i = -4\pi \lambda T\Tb$   and the flow equation for energy eigenvalues, can be readily identified as components of the Einstein equations for pure AdS$_3$ gravity.  A powerful fact is that these results on the QFT side are universal, applying to any deformed CFT and any energy eigenvalue of such a theory (at least any eigenvalue that evolves smoothly from the original CFT).

However, typical QFT states map not to pure gravity configurations in the bulk but rather to solutions with nontrivial matter fields present.   The presence of matter fields requires one to modify the CFT beyond just deforming it by $T\Tb$.  In particular, the full deformation must be manifestly non-local in position space in order to reproduce the momentum-space non-analyticity noted below \sj. This may be disappointing, but was nevertheless to be expected from e.g the well known failures of boundary-correlator micro-locality and strong subadditivity of entropy \HH\ known to occur when the bulk dual is subjected to a strict radial cutoff and is precisely what was argued in \HeemskerkHK\ and \FaulknerJY.

We also considered stress tensor correlation functions, focussing on two and three point functions computed to leading nontrivial order in the deformation parameter $\lambda$.  Here we found agreement, provided one compares results at the same order in $1/c$ perturbation theory, taking into account that $\lambda \sim 1/c$. An obvious task for the future is to extend this statement to all correlators computed at the classical level in the bulk.  The deformed CFT also makes definite predictions for bulk correlators at loop level, and verifying these would be interesting, as it involves the novel question of quantum gravity effects in a space with a boundary at a finite location.  One could also consider mixed correlators involving both the stress tensor and scalar operators, and the associated question of what new operators need to be added to the QFT to match bulk results.

Finally, we should note that in \McGoughLOL\ the authors have discussed defining a deformed CFT via an alternate Hubbard-Stratanovich type construction, and it would be useful to understand how correlators computed in that theory are related to those studied here.

\vskip .2cm
\noindent
{\bf Acknowledgements}
\vskip .1cm
  We thank David Berenstein, Mert Besken, Thomas Dumitrescu, Ashwin Hegde,  Ying-Hsuan Lin and Mark Mezei for useful discussions.
  PK is supported in part by NSF grant PHY-1313986. JL acknowledges support from the U.S. Department of Energy, Office of Science, Office of High Energy Physics, under Award Number DE-SC0011632. The work of DM was supported in part by the U.S. National Science Foundation under grant number PHY15-04541 and also by the University of California.

\listrefs
\end